\documentclass[letterpaper]{article} % DO NOT CHANGE THIS
\usepackage{aaai2027}  % DO NOT CHANGE THIS
\usepackage[hyphens]{url}  % DO NOT CHANGE THIS
\usepackage{graphicx} % DO NOT CHANGE THIS
\usepackage{natbib}  % DO NOT CHANGE THIS AND DO NOT ADD ANY OPTIONS TO IT
\usepackage{caption} % DO NOT CHANGE THIS AND DO NOT ADD ANY OPTIONS TO IT
\usepackage{algorithm}
\usepackage{algorithmic}
\usepackage{amsmath}
\usepackage{newfloat}
\usepackage{listings}
\DeclareCaptionStyle{ruled}{labelfont=normalfont,labelsep=colon,strut=off} % DO NOT CHANGE THIS
\floatstyle{ruled}
\newfloat{listing}{tb}{lst}{}
\floatname{listing}{Listing}

\usepackage{booktabs}

\title{Tunneling the Loss Landscape: Bypassing Memorization with Monte Carlo Parameter Swapping}

\author {
  Lai Shun Chan\textsuperscript{\rm 1},
  Xiaotian Zhang\textsuperscript{\rm 1},
  Yue Shang\textsuperscript{\rm 1,\rm 2},
  Ge Zhang\textsuperscript{\rm 1}\corresponding,
  Entao Yang\textsuperscript{\rm 3}\corresponding
  \\ 
}
\affiliations {
    \textsuperscript{\rm 1}Department of Physics, City University of Hong Kong, Hong Kong, China\\
    \textsuperscript{\rm 2}Department of Physics and Astronomy, University of Pennsylvania, Philadelphia, PA, USA\\
    \textsuperscript{\rm 3}Innovation Campus Delaware, Air Liquide, Newark, DE, USA\\
    gzhang37@cityu.edu.hk, entao.yang@airliquide.com
}

\begin{document}

\maketitle

\begin{abstract}
Grokking is a striking phenomenon in neural network training, where a model can undergo a prolonged period of pure memorization before abrupt generalization. 
While previous works have attempted to interpret it through classical machine learning mechanisms like weight norm, recent research draws an analogy from statistical physics, framing grokking as a form of computational glass relaxation.
This theory defines the initial memorization as a result of `fast cooling' where the training loss is reduced so quickly that a glass state is formed, followed by a `slow relaxation' towards final generalization.
Although providing a unifying framework for representative grokking theories, this perspective has remained largely at the theoretical on macroscopic level without direct empirical validation on training dynamics.
Here we introduce a three-component framework to directly characterize the training dynamics via parameter mobility (PM), and two representative measurements from glassy dynamics: replica correlation (RC) and fractal dimension (FD).
We demonstrate that standard optimization presents clear signatures of glass dynamics and inherently traps the grokking network in a kinetic arrested memorization state with a collapsed mobility, strong history dependence, and channel-like motions.
This quantitative agreement motivates us to introduce State-Aware Monte Carlo Parameter Swapping (SAM-Swap), an optimization plug-in that can accelerate generalization, inspired by swap Monte Carlo algorithm widely used in glass dynamics. 
Comparing SAM-Swap, weight decay, and Gaussian gradient noise, we find that accelerated generalization is consistently associated with random exploration in the parameter space, similar to diffusion in physics.  
\end{abstract}

\section{Introduction}
%While modern overparameterized neural networks can memorize even randomly labeled finite training sets, they usually generalize remarkably well in unseen data, making generalization one of the central problems in AI research \cite{zhang2016understanding}. 
Modern overparameterized neural networks (NNs) can perfectly memorize even randomly labeled finite training sets, yet they often generalize with surprising accuracy to unseen test data \cite{zhang2016understanding}.
Explaining this apparent contradiction remains one of the most fundamental open problems in AI \cite{zhang2016understanding,Belkin2019}.  
%Grokking provides a particularly surprising example of the generalization: a model first achieves near-perfect training performance while remaining at chance-level accuracy on test data (memorization), and then abruptly generalizes after a prolonged training.
Grokking exemplifies this paradox: models achieve near‑perfect training accuracy while remaining at chance‑level on test data (memorization), then abruptly generalize after prolonged training.
First observed in transformer-based models trained on algorithmic datasets \cite{power2022grokking}, grokking has later been reported across a broad range of architectures and domains including vision, language, and molecular learning tasks \cite{liu2022omnigrok, humayun2024grokking}.
%The pronounced timescale separation between memorization and generalization makes grokking a controlled testbed for distinguishing competing theories of neural-network generalization, thereby making a mechanistic understanding of grokking important to the broader study of how neural networks generalize. 
The pronounced timescale separation between memorization and generalization makes grokking a useful testbed for mechanistic theories of generalization.

%------------talk about how physics connectes with neural network-----------------
Existing theories of grokking can be broadly organized by the level where they characterize the memorization-to-generalization transition. 
Weight norm and landscape-based theories describe the process macroscopically, relating the delayed generalization to the evolution of the overall parameter norm through distinct training and test loss regimes, where test performance is favorable only within a narrower “Goldilocks” zone \cite{liu2022omnigrok}.
Circuit-based explanations offer a possible microscopic refinement of this picture, describing competitions between rapidly learned but parameter-inefficient memorizing computations and slowly formed, norm-efficient generalizing circuits \cite{nanda2023progress, merrill2023tale, varma2023explaining}. 
Representation-level theories instead emphasize the delayed feature learning and the gradual alignment of hidden representations with the underlying task structure \cite{kumar2023grokking, liu2022towards}.
Optimization-centered studies further investigate the dynamics driving these changes, including the instability of adaptive optimizers \cite{thilak2022slingshot}, slowly varying gradients \cite{lee2024grokfast}, effective learning rates \cite{prieto2025grokking}, and noise \cite{ersoy2026noise}. 

While these explanations illuminate different perspectives of grokking and are not mutually exclusive, they are formulated primarily in terms of machine learning-specific quantities.
A recent study introduced a complementary physics-inspired perspective and proposed that grokking can be interpreted as a form of \textit{computational glass relaxation}, which slowly evolves to a final equilibrium state \cite{zhang2025grokkingglass}. 
The key question is whether delayed generalization occurs because the generalizing solutions are intrinsically rare or separated from memorizing solutions by hard-to-access regimes of parameter space.
By conceptualizing each parameter configuration as a possible state of a statistical-mechanical system, training loss becomes its effective energy, and the Boltzmann entropy measures the size of solutions that exist at different levels of training loss and test accuracy.
%The key question is whether delayed generalization occurs because generalizing solutions are intrinsically rare or separated from memorizing ones by hard-to-access regions of parameter space. Viewing each parameter configuration as a state in a statistical-mechanical system, training loss then acts as effective energy while Boltzmann entropy measures the volume of solutions at different loss and accuracy levels.
Results show that generalization states occupy a much larger solution space than the memorization one, which also supports the \textit{volume hypothesis} \cite{chiang2023loss, peleg2024bias, pakman2026revisiting}.
Moreover, the number of accessible states does not collapse along the transition between memorization and generalization, while the training loss also decreases monotonically \cite{zhang2025grokkingglass}. 

In physical terms, these suggest the absence of both macroscopic entropic and energetic barriers, respectively, making grokking a process free of macroscopic thermodynamic barriers.
However, typical training trajectories via AdamW remain in a low-entropy regime with poor generalization for a long period, despite the transition to high‑entropy generalizing states being both accessible and statistically favored. 
This motivates the interpretation of grokking as computational glass relaxation to an equilibrium state with max entropy.
%In the context of physics, when an accessible and statistically favored transition is slow, the system is generally kinetic arrested. 
%On the one hand, the absence of macroscopic entropic and energetic barriers indicates that grokking is free of thermodynamic obstacles. On the other hand, training trajectories under AdamW often remain kinetically arrested in low‑entropy regimes with poor generalization for extended period,despite the transition to high‑entropy generalizing states being both accessible and statistically favored. From a physics perspective, such delayed relaxation is characteristic of glassy dynamics in condensed matter systems. This motivates the conclusion that grokking should be understood as computational glass relaxation.

While entropy analysis established the macroscopic accessibility of generalizing solutions, it does not characterize the dynamics directly but simply attributes grokking to slow parameter mobility caused by weak gradients after rapid loss reduction.
Here, we address this gap through a three-component framework combining parameter mobility (PM) with two representative measures from glassy dynamics: replica correlation (RC) \cite{Berthier2011,Mezard2012} and trajectory fractal dimension (FD) \cite{Datseris2023}.
%PM measures the magnitude of stepwise motion during training. 
%RC is instead an inter-trajectory observable that measures the similarity between independently evolving training replicas branched from a same model state. 
%FD is an intra-trajectory observable that characterizes whether each optimization path remains persistent and channel-like orbecomes more tortuous. 
PM quantifies the magnitude of stepwise parameter motion during training. RC measures inter‑trajectory similarity between independently evolving replicas branched from the same initial state. FD characterizes intra‑trajectory dynamics, indicating whether optimization paths remain persistent and channel‑like or become tortuous.
%Thus, the three components respectively ask whether the network changes, how each individual trajectory changes, and whether the same state can evolve into alternative paths.
Thus, the three components respectively ask i) whether the network changes, ii) whether the same state can evolve into alternative paths, and iii) how tortuous each trajectory is.

With this framework, we find that for AdamW-optimized model, PM collapses once the network reaches memorization, despite the poor test performance. 
RC exhibits clear two-step relaxation coinciding with the memorization period and progressively slower decorrelation at longer waiting times, revealing strong history dependence and aging, which is a representative sign of glassy dynamics \cite{kob1995testing, kob1997aging, Li2026}.
%This together provides a clear dynamical diagnosis for grokking, supporting the interpretation as glass-like kinetic arrest in parameter space.
%RC exhibits clear two-step relaxation with plateau of the curve matching the grokking period. Further test reveal progressively slower decorrelation at longer waiting times. Together these signal reveals a strong history dependence and aging, which is a representative sign of glassy dynamics.
Meanwhile, FD decays to approximately ${1}$ after memorization, suggesting the slowed parameter motion becomes persistent and channel-like.
This together provides a clear dynamical diagnosis for grokking, supporting the interpretation of grokking as glass-like kinetic arrest in parameter space.

%To further test our kinetic arrest theory, the next question we ask is whether grokking can be accelerated with disruption. 
To further test our kinetic arrest theory, we asked: if grokking mimics glassy dynamics, could interventions employed in glassy systems research also shorten the generalization delay?
%Inspired by the swap Monte Carlo method which is widely used to accelerate equilibration in glass-forming systems by switching particle identities \cite{shiraishi2024swapmc}, we introduce State-Aware Monte Carlo Parameter Swapping (SAM-Swap), which randomly exchanges parameter values within each layer.
This motivates the development of State‑Aware Monte Carlo Parameter Swapping (SAM‑Swap), inspired by the swap Monte Carlo method widely used to accelerate equilibration in glass‑forming systems by switching particle identities \cite{Grigera2001,Berthier2016}. 
Following the same concept, SAM-Swap treats each parameter in NNs as a 'particle' and randomly exchanges their values within each layer.

%Surprisingly, we find this method can substantially shorten the generalization delay. 
%This is counterintuitive as trained parameters are generally assumed to encode an organized computation \cite{ainsworth2022git} where randomly exchanging their values would destroy the learned representations and lead to degrade performance. 
At first glance, such swaps might appear to destroy learned features and obstruct representation learning, as trained parameters are generally assumed to encode an organized computation \cite{ainsworth2022git}
Yet, surprisingly, SAM‑Swap substantially shortens the generalization delay. Analysis via our three‑component framework further shows that SAM‑Swap can restore PM, reduce RC, and increase FD, indicating a transition from arrested dynamics to accelerated relaxation toward equilibrium.
We also compare this structured intervention with large‑amplitude additive Gaussian noise and find it can produce a similar acceleration, showing how different forms of disruption alter the dynamical route to generalization.
%Additional experiments show that large‑amplitude additive Gaussian noise produces a similar acceleration, although both interventions alter training dynamics in ways that differ qualitatively from standard Adam and AdamW.

In this work, we extend the emerging connection between grokking and glassy dynamics to direct characterization of training dynamics. Our main contributions are:
\begin{itemize}
    \item \textbf{Three-component framework for grokking dynamics:}
    We introduced a unified framework combining parameter mobility (PM), replica correlation (RC), and trajectory fractal dimension (FD) to characterize training dynamics. Our experiments suggest that long memorization is due to collapsed mobility, history-dependency, and channel-like trajectories, which provides direct dynamical evidence for interpreting grokking as glass-like kinetic arrest.
    
    \item \textbf{State-aware swap-based intervention:}  
    Inspired by the swap Monte Carlo equilibration method in glass physics, we propose a swap-based perturbation triggered by low FD. We find it can substantially reduce the grokking time, indicating counterintuitive benefits of disruption.    
    
    \item \textbf{Comparison of perturbation strategies:}
    We systematically compare multiple perturbation methods, showing how they alter training dynamics and thus affect the grokking time from the microscopic level.
\end{itemize}
Together, our results support the view of grokking as a shift between dynamically distinct training regimes, and suggest that stability- (RC) and geometry-based (FD) observables can inform targeted interventions for accelerating generalization.

\section{Methods}
%In this experiment, we aim to establish replica correlation and path fractal dimension as physical diagnostics for neural network training dynamics. Using a 1-layer transformer trained on the $(x^2 + y)mod67$ arithmetic task \cite{zhang2025grokkingglass}, we track the evolution of these metrics throughout training under standard cross-entropy optimization with AdamW. To assess the sensitivity of the optimization trajectory to local parameter perturbations, we apply small Gaussian perturbations to the parameters every epoch to probe the parameter space along the training trajectory. The result are drawn from the average of 20 models trained over 10000 epoch $t$. 
Here we follow the previously reported setup in the original glass theory work and experiment on a 1-layer transformer on the arithmetic task of $(x^2 + y) \bmod 67$ \cite{zhang2025grokkingglass}.
Results are averaged over 20 independent runs, each trained for 10,000 epochs($t$). 
%We employ RC and FD as physical diagnostics of neural network training dynamics. Experiments uses a 1-layer transformer trained on the $(x^2 + y)mod67$ arithmetic task \cite{zhang2025grokkingglass}, we track the evolution of these metrics throughout training under standard cross-entropy optimization with AdamW. Results are averaged over 20 independent runs, each trained for 10,000 epochs($t$). 
All models use the same training dataset, with 30\% of the available samples reserved for evaluating test accuracy. For ease of presentation, we define the onset of generalization as the point where the test accuracy first reaches 99\%. Unless otherwise stated, all references to grokking time correspond to the period after training accuracy reaches 99\%, and before test accuracy reaches 99\%.

\subsection{Replica-Correlation Functions}
%To quantify the similarity between independently evolving training trajectories, we define a replica-correlation function based on the Pearson correlation between model parameters.
To quantify the similarity between independently evolving training trajectories, we define a replica-correlation (RC) function based on the Pearson correlation between model parameters. We first train a model up to a waiting time $t_{\rm w}$, after which independent training processes (replicas) are spawned and evolved separately under perturbations. The correlation of pairs of replicas at each $t$ is measured.

Let $\mathbf{S}_l^{a}(t)$ denote the flattened parameter vector of layer $l$ for replica $a$ at training time $t$. We define the replica-correlation function as
\begin{equation}
q_l^{ab}(t,t_{\rm w}) = \mathrm{corr}\big(\mathbf{S}_l^{a}(t+t_{\rm w}), \mathbf{S}_l^{b}(t+t_{\rm w})\big),
\label{eq:rc}
\end{equation}
where $\mathrm{corr}(\cdot,\cdot)$ denotes the Pearson correlation coefficient computed over all parameters in layer $l$. This ensures that the correlation reflects the scale-invariant alignment between parameter configurations. Here, $a,b = 1,2,\ldots$ index different replicas. 
%Each pair of replicas is initialized from the same model state: the network is first trained up to a waiting time $t_{\rm w}$, after which independent training processes (replicas) are spawned and evolved separately under stochastic optimization and perturbations. The variable $t$ measures the elapsed epoch since branching.
%In practice, for each layer, parameters are flattened into a single vector and the Pearson correlation is computed pairwise across all replica pairs. 
The reported correlation is averaged over all replica pairs, layers, and independent runs. 
%Although different layers exhibit correlations with slightly varying magnitudes, we find that their temporal evolution follows a consistent qualitative trend, justifying averaging to extract the global behavior of the model.
High correlation indicates that nearby trajectories remain stable under perturbations, whereas a reduction in correlation signals that perturbations can lead to divergent evolution.

\subsection{Fractal Dimension of the Training Trajectory}
To characterize the geometric structure of neural network training dynamics, we estimate the fractal dimension (FD) of the parameter trajectory in weight space. Let $\mathbf{R}_t$ denote the flattened vector of all model parameters at training step $t$. We define the one-step increment as
\[
\Delta \mathbf{R}_t = \mathbf{R}_{t+1} - \mathbf{R}_t,
\]
and the cumulative arc length of the trajectory up to time $t$ as
\[
s_t = \sum_{t' \le t} \|\Delta \mathbf{R}_{t'}\|_2.
\]

For a given time lag $\tau$, we define the lag-dependent quantities
\[
\Delta s(t,\tau) = s_{t+\tau} - s_t, \qquad 
\Delta R^2(t,\tau) = \|\mathbf{R}_{t+\tau} - \mathbf{R}_t\|_2^2.
\]
We evaluate $\tau$ in the range of $\tau=1,....[t/2]$, ensuring that each lag fits entirely within the fractal window size $(T)$. To reduce sensitivity to non-stationarity and large fluctuations in parameter updates, we aggregate these quantities using the median, obtaining $\Delta s(\tau)$ and $\Delta R^2(\tau)$. We then assume a scaling relation of the form
\[
\Delta R^2(\tau) \propto \Delta s(\tau)^{\alpha},
\]
and estimate the exponent $\alpha$ via linear regression in log-log scale over a selected range of $\tau$,
%To probe temporal variations in the geometry of the trajectory, we perform this estimation within a sliding window of 50 training epochs, 
yielding a time-dependent exponent $\alpha(t)$ and the corresponding FD
\[
D_f(t) = \frac{2}{\alpha(t)}.
\]
The fractal dimension quantifies the relation between the accumulated path length and the net displacement of the trajectory. Values of $D_f \approx 1$ indicate that the trajectory evolves in a relatively straight or ballistic manner, where displacement scales proportionally with path length. In contrast, larger values of $D_f$ correspond to increasingly indirect or tortuous motion, where the trajectory covers more path length relative to its net displacement.

In our experiments, $D_f$ is estimated within a sliding window of $T=20$ epochs to capture local changes in trajectory geometry. We choose this value empirically, so that the window is sufficiently short to resolve the dynamical features emerging on the $10^2$ epoch timescale (beginning of memorization), yet long enough to suppress excessive fluctuations in the FD estimate. 
%We also considered variable window sizes, but found that FD estimation is highly sensitive to the selected window length. As a result, changing the window size during training introduced substantial variation in the estimated FD, making the resulting dynamics difficult to interpret. We therefore use a fixed 20-epoch window throughout the paper.

\subsection{Fractal Window Displacement}
To quantify parameter mobility (PM) within a training window, we measure the \textit{total displacement} $(dR)$ of each layer’s weights. For a given window, this displacement is defined as the Euclidean distance between the parameter vector at the end of the window and the parameter vector at its beginning:
\[
dR = \frac{1}{|l|} \sum_{l}\left\|\mathbf{R}_{t+T} - \mathbf{R}_t \right\|_2 ,
\]
This measure captures the net change in parameter space over the interval of the FD estimation window, rather than the cumulative path length. For network‑level analysis, the displacement values are averaged across all trainable layers.

\subsection{Small Additive Gaussian Noise as Probe}
%Under full‑batch Adam or AdamW without dropout layer, training is deterministic: replicas branched from the same state remain identical with $RC$ at $1$, and the trajectory is smooth so FD stays trivially $1$. To probe local trajectory and training dynamic, we inject weak Gaussian noise into all parameters, with strength set by the standard deviation $\sigma$. For Adam and AdamW we use $\sigma=10^{-5}$, which leaves the trajectory essentially unchanged while enabling meaningful measurement of RC and FD. All model probed this way will be mark as probed.
We use full‑batch Adam or AdamW without dropout to eliminate intrinsic sources of randomness, so that the noise level in the system is explicitly controlled. Replicas branched from the same state remain identical with $RC=1$, and FD is trivially $1$ in the absence of perturbations. 
Above this deterministic baseline, we inject zero‑mean Gaussian noise into all parameters with $\sigma=10^{-5}$. The value is chosen from control experiments where injecting the noise show no change in macroscopic training performance, yet enabling meaningful evaluation of RC and FD. Models subjected to this protocol are marked as probed.

\section{Observation}
\subsection{Parameter Mobility and Slow Relaxation}
As shown in Figure.~\ref{fig:avg_age}b, PM exhibits a clear transition from learning to memorization: $dR$ drops by orders of magnitude, from $\sim 1$ to $\sim 10^{-2}$ between $t \approx 0-10^2$, and continues to decrease toward $\sim 10^{-4}$ by the end of training. This progressive stagnation explains the slow learning dynamics, with mobility becoming vanishingly small around $t \approx 10^4$. Such behavior parallels slow relaxation in glassy systems, where dynamics progressively freeze and escape from the arrested state becomes increasingly difficult. The PM curves for different waiting times largely overlap, except for the initial datapoint, which reflects a larger first update due to untracked stochastic effects, while subsequent updates remain stable.

\subsection{Two-step Relaxation of Replica Correlation}
As shown in Figure.~\ref{fig:avg_age}c, RC exhibits a characteristic two-stage behavior for the model with $t_w=0$. 
%During the initial phase of training, RC remains high, indicating that no alternative optimization path is found under perturbation. 
During the initial phase of training, RC begins to decorrelate immediately, as expected since replicas are initialized with different random perturbations and therefore follow distinct optimization paths.
As training progresses and test accuracy diverges from training accuracy (Figure~\ref{fig:avg_age}a), marking the onset of grokking, RC settles into a plateau.
This plateau suggests that the trajectory is confined within a restricted region of parameter space and remains deterministic to small perturbations.
%Figure~\ref{fig:avg_age}, for the model with with $t_w=0$, shows that RC exhibits a characteristic two-stage behavior. During the initial phase of training, RC remains high, indicating that no alternative optimization path is found under perturbation. As training progresses and test accuracy diverges from training accuracy  (Figure~\ref{fig:avg_age}), signaling the onset of grokking, RC decorrelate and enter a plateaus. This plateau, despite the small drop, indicates that the trajectory is confined to a restricted region of parameter space and remains robust to small perturbations.
%At later times, coinciding with the onset of generalization, we observe a marked decrease in replica correlation for the perturbed model with $t_w=0$. 
At later times, coinciding with the onset of generalization, RC decreases markedly for the perturbed model, reflecting reduced inter‑trajectory similarity.
%Importantly, the baseline (unperturbed) training undergoes a similar evolution in training and test performance, yet its RC remains unchanged. 
%The observed decorrelation therefore cannot be attributed solely to the progression of training itself. 
%Rather, it indicates that the small perturbation is effective at separating nearby trajectories, suggesting a transition toward a more perturbation-sensitive and exploratory dynamical regime.
This indicates that the small perturbation have effectively separated replicas' trajectories, suggesting that there are alternative solutions accessible in the proximity at this time of the optimization path.
%This interpretation is further supported by the temporal alignment between the plateau in RC and and the collapse of the FD toward $D_f \approx 1$ (Figure~\ref{fig:avg_age}e). 
%Together with the FD results we presented below, indicate that the optimization trajectory becomes both perturbation-stable and geometrically constrained, following a comparatively direct path through parameter space during grokking, and transitions toward a more exploratory, perturbation-sensitive regime at generalization.
%suggesting that during grokking the optimization trajectory becomes perturbation‑stable and geometrically constrained, following a comparatively direct path through parameter space. 
%At generalization, however, perturbations reduce RC and increase FD, revealing additional solutions in the proximity of the trajectory and enabling broader exploration of parameter space.
Together these observations suggest that during the memorization phase the optimization trajectory remains constrained, with replicas confined to a small region of parameter space and converging to only a few similar solutions, whereas at generalization perturbations separate replicas and reveal nearby alternatives, enabling broader exploration of parameter space.

\subsection{Aging Behavior in Training Dynamics}
A signature of glass dynamics is observed in the dependence of replica correlation on the waiting time $t_w$, as shown in Figure~\ref{fig:avg_age}c. For replicas branched early ($t_w=0$), the correlation exhibits a two-step decay, consisting of an initial drop followed by a plateau and a subsequent relaxation, as we detailed above. 
In contrast, for larger waiting times (e.g. $t_w=10^2,10^3$), the initial decay is strongly suppressed, and the correlation remains close to $q \approx 1$ over an extended period before exhibiting only a minor decrease at later times.

This dependence on $t_w$ indicates that the system exhibits glass-like aging behavior: the response to perturbations depends on how long the system has already evolved. In particular, trajectories that have evolved longer become increasingly stable to perturbations and fail to exhibit rapid relaxation. The coexistence of two-step relaxation at $t_w=0$ and aging provides compelling evidence for glass-like dynamics, where the system enters a slow, history-dependent regime with constrained exploration \cite{Li2026}.
Within our framework, this behavior suggests the emergence of a confined, perturbation-stable phase during grokking, 
%The replica decorrelation at larger waiting time suggests that this glass-like regime is eventually destabilized as the system regains dynamical flexibility and resumes exploration of parameter space. 
which is eventually destabilized as the system regains dynamical flexibility and resumes exploration of parameter space.

\subsection{Dynamical Transition in Trajectory Geometry}
\begin{figure}[htbp]
    \centering
    \includegraphics[width=\columnwidth]{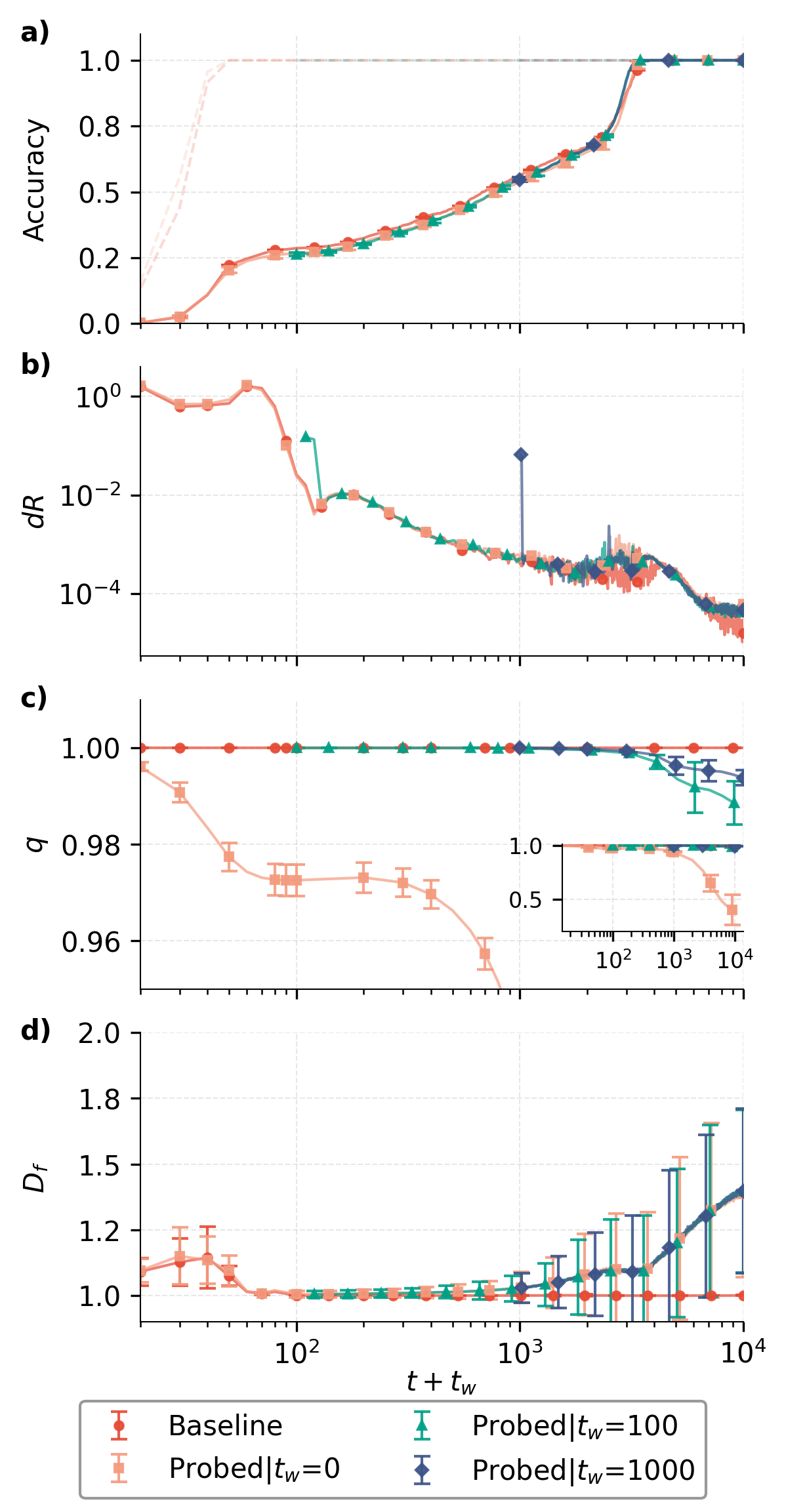}
    \caption{\textbf{Comparison across waiting times}. All data are from AdamW with learning rate $10^{-2}$ and weight decay $10^{-1}$, optimized for shortest grokking time. Baseline uses AdamW; probed model adds small Gaussian noise to test trajectory sensitivity. The x-axis is aligned to $t + t_w$. (a) Test accuracy (solid lines show Test accuracy; translucent line shows training accuracy).(b) Parameter mobility, error bars were omitted to declutter the plot and provide a clearer view of the decreasing $\Delta R$. Notable fluctuations are observed in the range from $0$ to $10^{2}$. (c) RC average across layers, with the insert showing the full range of replica decorrelation. (d) FD of Path average across layers. Occasionally, the regression yields FD estimates slightly below 1. This arises from fluctuations introduced by added noise, which perturb the slope in the log–log fit. As such, values below 1 are artifacts of estimation instability rather than evidence of sub‑linear trajectories.  
   }
    \label{fig:avg_age}
\end{figure}

As shown in Figure~\ref{fig:avg_age}d, the evolution of the FD exhibits a clear correspondence to different training stages. 
Initially, $D_f$ fluctuates moderately above 1, reflecting nontrivial exploration of parameter space as the model fits the training data, consistent with relatively high PM.
Once training accuracy reaches 99\%, $D_f$ collapses toward 1, indicating that the optimization trajectory becomes effectively low-dimensional and ballistic‑like. Even under perturbations, the model is not allowed to stray far from a narrow, channel‑like paths, reflecting strong confinement of the dynamics. This interpretation is further supported by the temporal alignment between the RC plateau and the collapse of $D_f$, consistent with restriction to a limited set of directions in parameter space and suppressed PM.
After a prolonged period of $D_f=1$, we observe a gradual increase in FD, signaling an expansion of the explored parameter space. This increase occurs on a similar timescale as the observed RC decrease.
At this time, the model has already generalized; this behavior then reflects the presence of multiple nearby minima. Perturbations induce diffusion in the proximity of the solutions, while the system remains strongly confined to a single basin with low PM, analogous to being trapped in a sharp minimum.

%The increase in FD under perturbations thus reflects the emergence of accessible directions in parameter space, rather than a fully autonomous transition in the unperturbed dynamics. 
The joint evolution of PM, RC and FD supports a two-stage dynamical picture: an initial relaxation followed by a confined, perturbation-stable regime, and a subsequent increase in sensitivity where perturbations can induce more exploratory dynamics. This supports the view that grokking corresponds to a phase in which the training trajectory is intrinsically low-dimensional, with stable and slow dynamics.
%while generalization is associated with an increasing susceptibility to perturbations that enables broader exploration of parameter space.
%While generalization is associated with an increasing susceptibility to perturbations that restore diffusive motion near minima and enable broader exploration of parameter space, it also reveals multiple nearby solutions that become accessible under perturbation.

\subsection{State-Aware Monte Carlo Parameter Swapping (SAM-Swap)}
%To further test our interpretation of grokking as a dynamically stable, low-dimensional regime, we introduce an intervention that perturbs the training trajectory based on the observed FD. 
To further examine our interpretation of grokking as a glass‑like dynamical phenomenon, we introduce an intervention designed to disrupt the training trajectory.
%The core hypothesis is that when the trajectory becomes strongly constrained, disrupting this stability should facilitate exploration and accelerate generalization.
Our core hypothesis is: disrupting constrained trajectories facilitates exploration and accelerates generalization, similar to real glassy systems \cite{Berthier2011}.
%Empirically, the onset of the grokking regime is consistently associated with a reduction in fractal dimension toward values close to 1, indicating effectively low-dimensional dynamics. 
Empirically, grokking onset coincides with FD dropping toward 1, indicating low‑dimensional dynamics.
%We therefore use a threshold on $D_f$ as a practical indicator of entry into this constrained regime. The precise value of the threshold is not critical provided it lies near the regime transition observed in practice. While moderate changes produce qualitatively similar behavior, extreme choices (e.g., $D_f \approx 2$) would no longer distinguish the constrained regime from normal training dynamics and would therefore lose their diagnostic value. In our experiment, we have chosen $df_{\text{glass}}=1.1$, the threshold is selected empirically to align with the observed onset of constrained dynamics.
We use a threshold on $D_f$ to indicate entry into the constrained regime, selected empirically as $d_{f.\text{glass}}=1.1$ to align with the observed onset of arrested dynamics. 
%Our experiment suggests the exact value is not critical as long as it lies near the regime transition; moderate changes yield similar behavior, whereas extreme choices (e.g., $D_f \approx 1.01$) fail to distinguish constrained dynamics from normal training and lose diagnostic value.
Our experiments indicate that the exact threshold value is not critical as long as $d_f > 1$; moderate adjustments yield comparable behavior, whereas extreme settings (e.g., $d_{f.\text{glass}} \approx 1.01$) produce similar overall trends but lead to highly fluctuating PM.

%When this condition is met, we apply a swap-based perturbation to the model parameters. Specifically, we introduce a structured, non-local perturbation by randomly exchanging pairs of parameter values within the same layer. 
When the FD threshold is reached, we apply a swap-based perturbation by randomly exchanging parameter pairs within the same layer. \cite{Grigera2001,Berthier2016}
%The amount of parameter swapped depended on the swap ratio $r$, which was varied from $10^{-1}$ to $10^{-5}$. We observe best performance with $r=10^{-2}$ in terms so stability of training and fast generalization, while smaller ratios progressively recover the baseline grokking dynamics.
%This indicates that only a minimum perturbation strength is required, with limited sensitivity to the precise value of $r$ above this threshold. 
%The amount of parameter swapped depended on the swap ratio $r$, varying swap ratio $r$ from $10^{-4}$ to $10^{-1}$, we find that $r=10^{-2}$ yields the most stable training and fastest generalization, while smaller ratios recover baseline grokking, indicating only a minimum perturbation strength is required.
The amount of parameters swapped depends on the swap ratio $r$. We find that $r=10^{-2}$ yields the most stable training and fastest generalization, while smaller ratios recover baseline grokking and larger ratios tend to induce excessive fluctuations. For details of the experiment regarding glass limit and swap ratio, please refer to the technical supplement.
Such an exchange operation preserves the overall distribution of parameters while redistributing the configuration in parameter space.
%To prevent excessively disruptive perturbations, we introduce a loss cap $L_{\text{cap}}$ as a safeguard. The loss cap is set to the initial training loss before optimization begins, and any perturbation that causes the training loss to exceed this value is rejected. This criterion occasionally filters out unusually destructive parameter updates while leaving the majority of proposed swaps unaffected. As such, it serves primarily to improve the robustness of the intervention rather than to shape the overall training dynamics. 
To prevent disruptive updates, we impose a loss cap $L_{\text{cap}}$ equal to the initial training loss, rejecting perturbations that lead to significantly large error. 
This safeguard improves robustness without altering overall dynamics.

We select FD as the trigger for swap interventions because alternatives are less practical: RC requires multiple replicas, so tracking it would be computationally expensive, and PM's magnitude depends heavily on the optimizer, initialization, and hyperparameters. 
In contrast, FD is naturally bounded between 1 and 2, making it straightforward to set a threshold near regime transition and providing a more robust, generalizable criterion for triggering swaps.

As shown in Figure~\ref{fig:all_noise}a, the swap intervention substantially reduces the grokking time, with models generalizing at approximately $t\approx650$ epochs compared to $t\approx3000$ epochs for the AdamW model.  
This result supports the view that the grokking regime corresponds to a dynamically stable phase in which the training trajectory is constrained and resistant to perturbations. 
The swap intervention instead disrupts this stability, allowing the optimization process to follow alternative trajectories that reach generalization more efficiently. More broadly, these results show that the timing of generalization depends not only on the structure of the loss landscape, but also on the specific trajectory taken during optimization, and how it is influenced by perturbations.
\begin{algorithm}[htbp]
\caption{State-Aware Swap Perturbation with Fractal Dimension Trigger}
\label{alg:swap}
\begin{algorithmic}[1]
\REQUIRE Model parameters $\theta$, training data $\mathcal{D}$, swap ratio $r$, glass threshold $df_{\text{glass}}$
\STATE Save training loss at $t=0$: $L_{\text{cap}}$
\FOR{each transformer layer $\ell$}
    \STATE Compute fractal dimension $df_\ell$
    \IF{$df_\ell < df_{\text{glass}}$}
        \STATE Store backup $\theta_{\text{backup}} \leftarrow \theta$
        \STATE Sample $2 \cdot \lfloor rN_\ell \rfloor$ random indices $\{i_k\}$ in layer $\ell$
        \FOR{$k = 1$ to $\lfloor rN_\ell \rfloor$}
            \STATE Swap $\mathbf{v}_\ell[i_{2k-1}]$ and $\mathbf{v}_\ell[i_{2k}]$
        \ENDFOR
        \STATE Compute new training loss $L_{\text{new}}$
        \IF{$L_{\text{cap}} < L_{\text{new}}$}
            \STATE $\theta \leftarrow \theta_{\text{backup}}$ \COMMENT{reject}
        \ENDIF
    \ENDIF
\ENDFOR
\STATE \textbf{return} updated $\theta$
\end{algorithmic}
\end{algorithm}

\subsection{Comparing perturbation methods}
Recent studies demonstrate that stochastic perturbations can strongly affect grokking dynamics, often accelerating or even eliminating delayed generalization \cite{agrawal2025grokking,ersoy2026noise}. 
To investigate how perturbations influence grokking dynamics, we consider four training settings: (i) Adam probe, (ii) AdamW with probe, (iii) Adam with large Gaussian noise, and (iv) Adam with the SAM-Swap introduced in the previous section. 

It is important to distinguish between perturbations used for measurement and those used as interventions. 
%Under full-batch Adam or AdamW optimization, training is effectively deterministic, so independently evolved replicas branched from the same model state remain identical and maintain $RC=1$. To probe the sensitivity of these trajectories to perturbations, we apply weak Gaussian noise to all parameters, with perturbation strength controlled by the standard deviation $\sigma$. For the Adam and AdamW measurements, we use $\sigma=10^{-5}$, which has negligible impact on the underlying training trajectory while enabling RC and FD to be measured meaningfully. 
%In contrast, for large noise we use $\sigma=10^{-2}$ as a genuine intervention intended to alter the optimization dynamics and promote escape from the constrained low-FD regime. 
Probing refers to the small Gaussian noise described in Methods. As a genuine intervention, we instead use larger noise applied in the same way but with $\sigma=10^{-2}$, deliberately altering the optimization dynamics to promote escape from the constrained low‑FD regime.
The proposed swap method serves a similar purpose, but induces structured non-local rearrangements of parameters rather than additive stochastic fluctuations. 

\begin{figure}[htbp]
    \centering
    \includegraphics[width=\columnwidth]{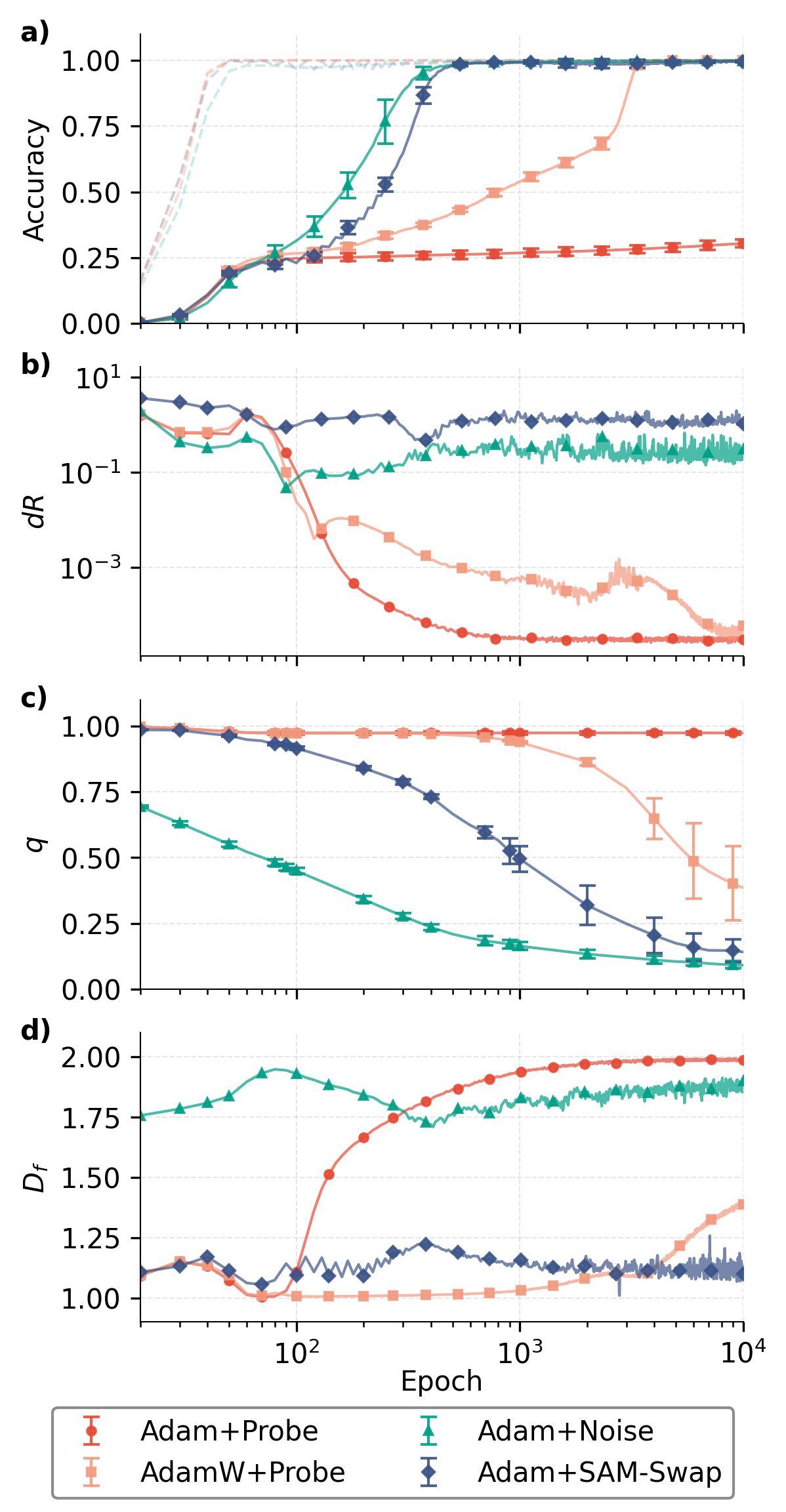}
    \caption{\textbf{Comparison across perturbation methods}: All models uses learning rate of $10^{-2}$, while AdamW additionally applies weight decay of $10^{-1}$. Both models are probed using a weak Gaussian probe ($\sigma=10^{-5}$), whereas Adam+Noise employs additive Gaussian perturbations with $\sigma=10^{-2}$ and Adam+Swap uses the proposed SAM-Swap. All hyperparameters are selected to minimize grokking time. (a) Test accuracy (solid lines show Test accuracy; translucent line shows training accuracy) b) Parameter Mobility (c) Replica Correlation from $t_w=0$. (d) Fractal dimension of path.  
    For fig (b) and (d), error bars were omitted in the plots to avoid excessive visual clutter. The Adam+Noise and Adam+Swap settings are inherently subject to large fluctuations due to the added perturbations; this variability is acknowledged here, and we instead present the average trend.}
    \label{fig:all_noise}

\end{figure}

%As shown in Figure \ref{fig:all_noise}, Adam with Weight Decay, Adam with large Gaussian noise, and Adam with swap intervention all reaches generalization earlier than the baseline full-batch Adam optimizer,
As shown in Figure \ref{fig:all_noise}a, training setting (ii),(iii) and (iv) all reaches generalization earlier than the baseline setting (i),
though with varying effectiveness. These results suggest that perturbations, regardless of form, can substantially alter the timing of delayed generalization.

\subsection{The three component framework}
Drawing together the evidence from PM, RC, and FD, we can now articulate a three‑component framework for understanding the dynamics of grokking. 
%We found that the prolonged memorization stage of grokking can be understood as a glass-like kinetic arrest in parameter space. The network is prevented from generalization as its dynamics become slow, history-dependent, and geometrically constrained.
Within this framework, the prolonged memorization stage of grokking emerges as a glass‑like kinetic arrest in parameter space: the network is prevented from generalization as its dynamics become slow, history‑dependent, and geometrically constrained. 
Although the generalizing solutions are abundant and accessible in principle \cite{zhang2025grokkingglass}, the optimization process has difficulty exploring alternative parameter configurations. 
%Such a glass-like kinetic arrest can be characterized via a three-component framework including mobility, history-dependence, and trajectory geometry.
This glass‑like arrest can be systematically characterized through the three components of mobility, history‑dependence, and trajectory geometry.

%Mobility measures the magnitude of parameter changes during training. As argued by previous work \cite{zhang2025grokkingglass}, in grokking systems the training loss is reduced too quickly, and so the gradients become small while parameters become stagnant. Our experiments also confirm this point as the parameters become nearly immobile after memorization, even though the test performance remains poor, as shown in fig.\ref{fig:all_noise}.
Mobility, measured by PM, quantifies the magnitude of parameter changes during training. Prior work \cite{zhang2025grokkingglass} shows that in grokking systems, the training loss reduces too quickly, gradients shrink while parameters stagnate. Our experiments also confirm this point as the parameters become nearly immobile after memorization, even though the test performance remains poor, as shown in Figure~\ref{fig:all_noise}b.
However, low mobility alone does not necessarily impose delay in generalization, just like slowing dynamics itself does not establish glass phenomena in  \cite{Dyre2006, Berthier2011}. 
A model could move slowly just because of some specific optimization settings, for example, a decaying learning rate. 
%Slow motion could have arised from other source, like a decaying learning rate.

%History dependence is a key feature of glass system in physics, which can be investigated with replica correlation.
History dependence, probed by RC, is a key feature of glass systems in physics \cite{Berthier2011,Mezard2012}. 
While our baseline dynamics show a clear two-step relaxation pattern, the slower decorrelation between replicas created after long waiting times further supports the analogy between grokking and glass relaxation. 
%This mimics the aging process in glass where the longer the network remains in the arrested state, the more difficult it escapes to an alternative trajectory.
This mimics glass aging: the longer the arrest, the harder escaping to an alternative trajectory becomes.
In other words, a high RC suggests that even when nearby copies of the model are given an opportunity to evolve independently, they continue to follow highly similar trajectories.
The optimizer therefore has limited access to alternative routes in parameter space.
%This distinguishes parameters' simple small update (low mobility) from genuine dynamical persistence.
When high RC coincides with low mobility, it indicates that slow dynamics are not merely small updates but may stem from limited alternative paths available to the optimizer, as shown in Figure~\ref{fig:all_noise}c.
Nevertheless, simply reducing the RC alone does not guarantee accelerated generalization. For an extreme example, RC can be simply reduced to zero by resetting all parameters after memorization, which destroys all the learned representations and cannot accelerate generalization.

%Trajectory geometry can be characterized via fractal dimension, which compares the parameters' net displacement and the total distance traveled during optimization. 
%We found that grokking networks fractal dimension decays to 1 after memorization, indicating that the parameters are following a straight, channel-like path. 
Trajectory geometry, measured by FD, compares net displacement to path length. During grokking, FD decays to 1 after memorization, indicating straight, channel‑like paths.
This suggests that the optimization follows a highly constrained motion: although the model could accumulate small updates, those updates cannot freely explore the parameter space.
It further explains why low mobility itself is insufficient, as two models could move by a similar total distance, with one following the same narrow direction while the other randomly explores many alternative directions before reaching the same level of generalization.
%This also highlights that two models may traverse the same total distance, or have the same PM, yet their dynamics can differ profoundly: one may follow a straight, channel‑like path confined to a narrow direction, while the other may take a tortuous trajectory that fully explores the surrounding landscape for alternative solutions before converging.
%A high FD itself is also insufficient, as shown in the Adam+Probe experiment in Figure \ref{fig:all_noise}, because it only means the trajectory is tortuous and does not tell whether the exploration remains locally or not. 
Additionally, FD alone is not sufficient, as shown in the Adam+Probe experiment in Figure\ref{fig:all_noise}d. 
Without considering PM or RC, a high FD only indicates diffusive motion, but does not tell whether the exploration remains local or not. 
%but this interpretation can be misleading. 
%For this case, the fact that replicas remain highly correlated suggests that exploration is local.

In summary, these three observables characterize complementary levels of the optimization dynamics. 
PM is a local, within-trajectory quantity that measures the magnitude of ongoing parameter motion.
FD is also evaluated within individual trajectories but characterizes their larger-scale geometry, distinguishing persistent channel-like motion from more tortuous exploration.
RC is instead an inter-trajectory quantity, which compares an ensemble of independently perturbed trajectories originating from the same network state and measures whether the system can access distinct future evolutions.
Their combination distinguishes general slow motion from a state where parameter update is simultaneously suppressed, geometrically channelized, and restricted from accessing alternative solutions.
Only when taken together do the three components provide a full picture: low PM may reveal that the system is not truly moving away from its initial solution despite apparent motion, while high RC indicates that replicas remain correlated, implying there are no alternative directions being explored; a high FD, on its own, simply reflects that the trajectory is tortuous, but without PM and RC it cannot distinguish whether such motion corresponds to genuine diffusion across minima or merely local wandering, where RC provide the answer for whether there are other minima to diffuse to. Therefore, PM, RC and FD jointly distinguish true diffusion and exploration of nearby solutions from motion that is geometrically complex but dynamically confined.
%Their combination distinguishes general slow motion from a state where parameter update is simultaneously suppressed, geometrically channelized, and restricted to alternative solution.

%By studying them in combination, we discover that grokking can be accelerated by random exploration of the parameter space, similar to diffusion in physics, as evident by increased mobility, decaying RC, and high FD. Our discovery that diffusion (random exploration) accelerates generalization is also consistent with previous work from the physics lens: diffusion causes an increase in entropy, and entropy is positively correlated with generalization in a variety of neural networks \cite{zhang2025grokkingglass, yang2026high}.
Our framework also explains why grokking can be accelerated by structured perturbations like SAM-Swap. 
Accelerating grokking via random exploration is similar to diffusion in physics, as evident by increased mobility, decaying RC, and high FD, which enables the system to escape kinetic arrest and thereby shortening the generalization delay.
This is also consistent with previous work from the physics lens: diffusion causes an increase in entropy, and entropy is positively correlated with generalization in a variety of neural networks \cite{zhang2025grokkingglass, yang2026high}.
%Interventions such as SAM‑Swap or large additive Gaussian noise restore diffusive motion where it is otherwise restricted by the glass‑like nature of grokking, reduce replica correlation, and expand the accessible parameter space. In this view, perturbations reveal multiple nearby solutions and enable the system to escape kinetic arrest, thereby shortening the generalization delay. More broadly, our discovery that diffusion accelerates generalization is consistent with the physics lens: random exploration increases entropy, and entropy has been shown to correlate positively with generalization in a variety of neural networks \cite{zhang2025grokkingglass, yang2026high}.

\section{Conclusion and Future Directions}
Overall, our findings suggest that the timing of generalization is shaped not only by which solutions exist in the loss landscape, but also by which regions of parameter space the training dynamics can access. This perspective reframes perturbations from generic noise sources into possible control mechanisms for modifying neural training trajectories. By jointly measuring mobility (PM), stability (RC) and geometry (FD), we provided a practical framework for studying when optimization becomes dynamically constrained and when it becomes capable of exploratory searching and ultimately reaching generalizable solutions.

Our experiments focus on modular arithmetic grokking and employ a local FD estimator based on a fixed temporal window. Future work should therefore address both methodological and empirical extensions. On the methodological side, more robust multiscale estimators are needed to determine whether the observed low-dimensional regime persists across timescales and window sizes. On the empirical side, it remains unclear whether FD-triggered perturbations generalize beyond modular arithmetic to other grokking tasks, overparameterized memorization settings, or realistic vision and language models. 

More broadly, dynamical observables such as FD, RC, or even curvature \cite{Keskar2017,Chaudhari2017,Sagun2017}, and representation stability \cite{Bietti2019,Mairal2016} may enable state-dependent adaptive training interventions that respond to the evolving optimization trajectory rather than applying perturbations uniformly throughout training.

\section{Limitation}
%\subsection{Timescale Sensitivity of Fractal-Dimension Estimation}
%A further limitation of our current analysis concerns the timescale over which the fractal dimension is estimated. In our experiments, the sliding-window FD is primarily computed using a relatively short window of 20 epochs. This choice was motivated empirically: although we also tested longer windows up to approximately 200 epochs, the shorter window produced the clearest correspondence with the early dynamical changes occurring between roughly $10^1$ and $10^2$ epochs. Longer windows tend to smooth over these transient dynamics and therefore obscure the initial collapse of the trajectory into the low-dimensional regime. 
Our main limitation is the timescale of FD estimation. We compute FD with a 20‑epoch sliding window, as longer windows (we tested up to 500 epochs) obscure early dynamics; 20 epochs best captured dynamics around $10^1- 10^2$ epochs. Longer windows smooth transients and hide the collapse into low‑dimensional regimes. Results of this test are presented in the technical supplement. 
%As a result, the reported FD should be interpreted as a local, short-timescale geometric diagnostic rather than a complete multiscale characterization of the training trajectory.
As a result, the reported FD should be interpreted as a local diagnostic, not a full multiscale characterization. 

Another weakness of the current FD design is the reliance on a fixed temporal window to describe what is inherently a multiscale parameter. This means dynamics with persistence times shorter or longer than the chosen window are either averaged out or missed.

Extending this analysis to larger temporal windows or multiple windows simultaneously presents both computational and methodological challenges. Computing FD over long timescales across many replicas, perturbation settings, and training runs is expensive. A natural workaround would be to estimate FD using representative sampling of the trajectory. 
However, we found experimentally that the resulting FD estimates are sensitive to both the sampling range and sampling density, since the method relies on a log--log regression over displacement and path-length statistics. Consequently, different choices of sampled lags or trajectory segments can lead to noticeably different estimated exponents. 
%This sensitivity suggests that while FD is a useful probe of training geometry, its current implementation requires further refinement before it can be treated as a robust, scale-independent order parameter for neural training dynamics.
This sensitivity suggests that while FD is a valuable probe of training geometry, its current formulation remains window‑dependent and requires multiscale refinement before it can serve as a robust order parameter.

%\section{Future Direction}
%Our experiments focus on modular arithmetic grokking and employ a local FD estimator based on a fixed temporal window. Future work should therefore address both methodological and empirical extensions. On the methodological side, more robust multiscale estimators are needed to determine whether the observed low-dimensional regime persists across timescales and model sizes. On the empirical side, it remains unclear whether FD-triggered perturbations generalize beyond modular arithmetic to other grokking tasks, overparameterized memorization settings, or realistic vision and language models. 

%More broadly, dynamical observables such as FD, replica correlation, curvature, or representation stability may enable state-dependent training interventions that respond to the evolving optimization trajectory rather than applying perturbations uniformly throughout training.

\section{Ethics Statement}
%This work is primarily a methodological and study of training dynamics in controlled algorithmic task. Its intended contribution is to improve empirical understanding of memorization, delayed generalization, and perturbation sensitivity in neural networks by introducing dynamical probes such as fractal dimension and replica correlation. The work aims to make aspects of neural optimization more measurable and interpretable.
%This work is a methodological study of training dynamics in controlled algorithmic task. It aims to improve understanding of memorization, delayed generalization, and perturbation sensitivity. The goal is to make neural optimization more measurable and interpretable.
This work is a methodological study of training dynamics in controlled algorithmic tasks, aimed at understanding memorization, delayed generalization, and perturbation sensitivity.
%The main ethical and methodological risk is not associated with the specific experimental task, but with the interpretation of models trained under perturbative dynamics. A model trained with state-aware perturbations may reach similar test accuracy to a model trained with standard methods, but this does not imply that the two solutions are equivalent. Because perturbation changes optimization trajectory, the resulting model may differ in representation, robustness, calibration, sensitivity to distribution shift, or failure modes. Therefore, improved generalization speed alone should not be taken as evidence that the perturbed model is functionally identical to a conventionally trained model.
%The main ethical and methodological risk is not associated with the specific experimental task, but with the interpretation of models trained under perturbative dynamics. Perturbed models may match standard test accuracy but are not necessarily equivalent, as perturbations alter trajectories, so models may differ in representation, robustness, calibration, and so on. Therefore, faster generalization alone does not prove functional equivalence to conventional training.
The main risk lies in interpreting models trained under perturbative dynamics: they may reach similar test accuracy to conventional ones yet differ in representation, robustness, or calibration. In this paper, models are trained only to minimize grokking time for studying dynamics.
%Another limitation concerns interpretability. Fractal dimension and replica correlation are indirect empirical diagnostics of training dynamics. They should not be interpreted as guarantees of robustness, reliability, fairness, or safety. Overstating these observables could lead to misleading conclusions about model behavior, especially in larger or more realistic systems.
Another risk is interpretability: FD and RC are indirect diagnostics of dynamics. They are not guarantees of robustness, reliability, fairness, or safety. Overstating them may mislead conclusions, especially in larger or realistic systems.
%A further limitation is interpretability: FD and RC are indirect diagnostics.

\section{Acknowledgments}
The authors thank National Natural Science Foundation of China for supporting this research (Grant 12405043, G.Z.).
We also thank computational resources provided by Bridges-2 at Pittsburgh Supercomputing Center and DeltaAI at National Center for Supercomputing Applications through ACCESS allocation CIS230096 (E.Y.).

\bibliography{aaai2027}

%moved all supplment material to technical supplment.tex
\end{document}